# Capital Market Performance and Macroeconomic Dynamics in Nigeria

**Fapetu, Oladapo[#1], Ojo, Segun Michael[*2], Balogun, Adekunle Alexander[#3],
Asaolu, Adeoba Adepoju[#4]**

[#]*Department of Finance, Federal University Oye-Ekiti, Nigeria*
[*]*Department of Economics, Redeemers University, Ede, Nigeria*
[1]oladapo.fapetu@fuoye.edu.ng, [2]ojosegunm@yahoo.com, [3]adekunleabalogun@gmail.com,
[4]asaoluadepojupraise@gmail.com

**Citation:** Fapetu, O., Ojo, S. M., Balogun, A. A., & Asaolu, A. A. (2021). Capital market performance and macroeconomic dynamics in Nigeria. *Fuoye Journal of Finance and Contemporary Issue, 1*(1), 29-37.

Abstract
*The study examined the relationship between capital market performance and the macroeconomic dynamics in Nigeria, and it utilized secondary data spanning 1993 to 2020. The data was analyzed using vector error correction model (VECM) technology. The result revealed a significant long-run relationship between capital market performance and macroeconomic dynamics in Nigeria. We observed long-run causality running from the exchange rate, inflation, money supply, and unemployment rate to capital market performance indicator in Nigeria. The result supports the Arbitrage Pricing Theory (APT) proposition in the Nigerian context. The theory stipulates that the linear relationship between an assets's expected returns and the macroeconomic factors whose dynamics affect the asset's risk can forecast an asset's returns. In other words, the result of this study supports the proposition that the dynamics in the exchange rate, inflation, money supply, and unemployment rate influence the capital market performance. The study validates the recommendations of Arbitrage Pricing Theory (APT) in Nigeria.*

## I. INTRODUCTION

The quest to investigate the nexus between capital market performance and the macroeconomic dynamics in Nigeria is motivated by two significant factors. One, macroeconomic dynamics is a critical issue in any economy, particularly in developing countries. The economic outcomes of government fiscal and monetary policies through development plans, national budget, reforms, and program of agenda; partly depend on the macroeconomic environment. Implementation of economic policies and development plans run smoothly in an enabling macroeconomic environment.

Therefore, the unanticipated changes or movements in core macroeconomic variables like money supply, inflation rate, exchange rate, and unemployment rate are critical determinants of economic outcomes.

Two, capital market development is essential to economic stability and growth. The capital market is a significant facilitator of economic activities. Funding of financial undertakings like the national development plans, reform programs, business growth and expansion, public and private sector investment projects, and provision of infrastructural facilities all lean on the capital market. In other words, capital market development is sine qua non for economic stability and creating of enabling macroeconomic environment for sustainable growth and development.

Another issue of interest in this study is the aftermaths of the last three global crises on the Nigerian economy. In other words, Nigeria's economy bumped into three major economic problems lately. These include the global financial crisis of 2008, the oil price crisis of 2014 and the COVID 19 crisis of 2020.

However, the trends in economic development indicators like the human development index, cost of living, the standard of living, and the unemployment rate bear out that the cumulative effect of those crises on the economy is highly consequential. In Nigeria, the sectors that are most traumatized by those crises include the financial sector, the macroeconomic environment and a host of others. Hence, the need to investigate how the dynamics in the macroeconomic environment impact the capital market performance in Nigeria. The outcome of this study will shed light on the nexus between the macroeconomic environment and capital market performance in Nigeria.

The capital market is the centre for the transaction of long-term funds to facilitate long-term projects needed to actualize the desired growth and development in an economy (Shallu, 2014). The link between the capital market and the real sector is traced through the effect of asset prices on households and firms' balance sheets. At the secondary level, on consumption and investment (Ankargren, Bjellerup, & Shahnazarian (2017). Put differently, consumption and investment are drivers of economic growth. The capital market influences economic growth through its effect on consumption and investment. The development of the capital market can also generate employment and boost income, thereby stirs economic growth.

The capital market in Nigeria comprises the Securities and Exchange Commission (SEC), which is responsible for regulating the activities of the Nigerian stock market. At the same time, the Nigerian Stock Exchange (NSE) supervises the operations of the quoted market (as a self-regulatory organization). Figure 1 depicts the trend in market capitalization share of GDP in Nigeria from 1993 to 2020. The graph shows that there has been no occasion of sustained growth in market capitalization in Nigeria. The trend in market capitalization in Nigeria oscillated too frequently, and it has not experienced any significant increase since the global financial crisis of 2008. The graph shows an unprecedented rise in market capitalization growth in Nigeria from 2006 to 2007. The financial crisis of 2008 caused market capitalization to fall drastically, and it has not had any significant improvement since that time.

The trend exhibits a little sign of progress in 2013. The advancement in the previous years ran to a halt in 2014 over the oil price crisis, which again dragged the level of market capitalization performance down. This analysis shows that the financial sector is highly vulnerable to external shocks because the effects of those externally motivated crises are too noticeable on the movement in market capitalization in Nigeria.

Figure 1: Market Capitalization as Percentage of GDP

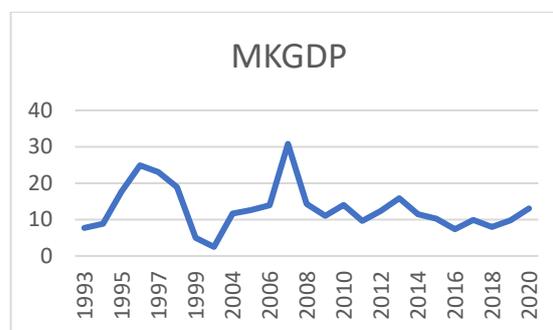

*Source:* Authors' compilation (2021)

In sum, the capital market is one of the key drivers of economic growth and development (Odetayo & Sajuyigbe, 2012; Okoye & Nwisienyi, 2013) because it plays a vital role in the mobilization of saving and investment to facilitate the production of goods and services in an economy (Okodua & Ewetan, 2013). The success of the capital market at facilitating growth and development through mobilization of savings and investment dependent on the macroeconomic environment in which a particular capital market apparatus is being operated. Therefore, this study seeks to examine the relationship between macroeocnomic dynamics and capital market performance in Nigeria.

II. LITERATURE REVIEW
Theoretical Framework
The Arbitrage Pricing Theory (APT) by Ross (1976) stipulates that asset returns are forecastable through the linear relationship between an asset's expected returns and the macroeconomic factors whose dynamics affect the asset's risk. This theory holds that while evaluating their alternative investment opportunities, investors usually factor in the effects of the risk-sensitive macroeconomic variables. In other words, some macroeconomic variables have a direct link with the risk that associate with investment and the risk premium. For instance, exchange rate, inflation, and money supply mean a lot to the investors because of their potential effects on the value of their assets and the returns.

Empirical Review
The empirical studies carried out on the relationship between capital market performance and macroeconomic issues like economic growth, exchange rate stability, inflation rate, employment generation, etc., in Nigeria and other developing economies include the study by Cynthia, Chinedum, & Ikechi (2021). This study examines the effect of capital market development on the economic growth in Nigeria and utilizes secondary data spanning 1983 to 2016. The data was analyzed using the autoregressive distributed lags model (ARDL). The results show that the number of listed securities has a positive and significant relationship with growth in Nigeria. In the study, the number of all share indexes is negative and significant. Enoruwa, Ezuem & Nwani (2019) examined the impact of the capital market on economic growth in Nigeria.

The study used secondary data from 1985 to 2015 using the ordinary least square (OLS) method. The study found that market capitalization, all share index, trade volume, and trade value exhibited a positive and significant relationship with economic growth in Nigeria. Edame & Okoro (2013) examined the effect of capital market performance on economic growth in Nigeria. The study utilized the ordinary least square (OLS) method. It observed a positive relationship between capital market performance and economic growth in Nigeria.

Alam and Hussein (2019) carried out a study titled impact of capital market on economic growth in Oman. The study utilized secondary data spanning 1960 to 2015. The data was analyzed using OLS. The results reveal a positive and significant relationship between capital market and economic growth in Oman. Anigbogu & Nduka (2014) examined the relationship between stock market performance and economic growth in Nigeria for the period 1987 to 2012. The study's data was analyzed using the vector error correction model (VECM) technique. The analysis results confirm the positive and significant relationship between stock market performance and economic growth in Nigeria. Acha & Akpan (2019) studied the relationship between capital market performance and economic growth in Nigeria. The study utilized secondary data spanning 1987 to 2014, and the data were analyzed using the vector autoregressive (VAR) method. The study observed a positive and significant relationship between capital market performance and economic growth in Nigeria.

The empirical studies on the nexus between capital market performance and the macroeconomic environment in Nigeria are massive. But few have decisively examined the plausibility of the theoretical propositions in the context of the Nigerian economy. For instance, the theories on capital market analysis and forecast include the Capital Asset Pricing Model (CAPM), Arbitrage Pricing Theory (APT), and a host of others. The empirical studies from developing countries, including Nigeria, usually make references to studies that evaluate those theories in developed countries. Not many have tested the applicability of those theories in developing economies. This issue had engendered a lot of inaccurate judgments and perceptions about economic matters in developing countries.

Therefore, this study examines the nexus between capital market performance and the macroeconomic dynamics in Nigeria in the light of the Arbitrage Pricing Theory (APT) framework. The study's outcome will clarify the relationship between the capital market performance and the macroeconomic dynamics in Nigeria and determine the applicability of the Arbitrage Pricing Theory (APT) in the context of the Nigerian economy.

## III. METHODOLOGY

The mathematical form of the arbitrage pricing theory is:

$$ER(Y) = \theta + \rho_1\gamma_1 + \rho_2\gamma_2 + \rho_3\gamma_3 + \cdots + \rho_n\gamma_n \quad\quad\quad\quad (1)$$

$ER(Y)$ is the expected return on asset, $\theta$- the riskless rate of return, $\rho_i$(Beta) is the asset's price sensitivity to factor, and $\gamma_n$- the risk premium associated with the factor. This study aims to determine the short and long-run coefficients that associate with the relationship between the returns on assets and the macroeconomic factors using the appropriate econometric technique in the spirit of the arbitrage pricing theory.

The conceptual framework depicted in figure 2 shows that return on assets is subject to the macroeconomic variables that have a direct or indirect influence on it in one way or the other. The risk-sensitive macroeconomic variables constitute the macroeconomic environment to the risk premium associated with the capital market securities. Investors usually give special consideration to possible changes in those macroeconomic variables while evaluating their alternative investment opportunities. The macroeconomic variables are not among the mainstream financial variables, but they are macroeconomic variables that influence the financial performance of the firms operating in the capital market.

**Figure 2: Conceptual Framework on the Nexus between Expected Return on Asset and Macroeconomic Variables**

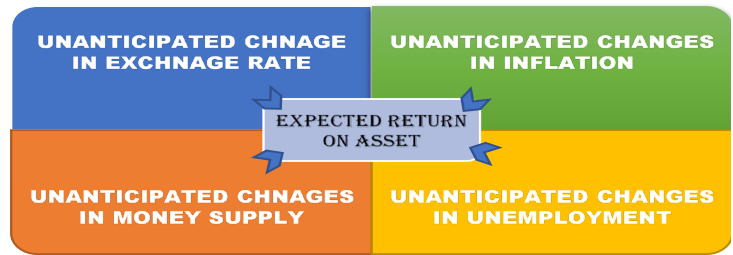

*Source:* Authors' compilation (2021)

### Model Specification

The dynamic relationship among the exchange rate, inflation, money supply, unemployment rate, and the expected return on assets is analysed in this study using vector error correction model (VECM) technique. The mathematical form of the model is stated as;

$$\Delta ROA_t = \alpha_1 + \sum_{j=1}^{n-1}\beta_j\Delta ROA + \sum_{j=1}^{n-1}\gamma_j\Delta INF_{t-j} + \sum_{j=1}^{n-1}\rho_j\Delta MSP_{t-j} + \sum_{j=1}^{n-1}S_j\Delta EXCH_{t-j} + \sum_{j=1}^{n-1}H_j\Delta UPY_{t-j} + \varphi_1 ecm_{t-1} + \varepsilon_{1t} \quad\quad (2)$$

$$\Delta INF_t = \alpha_1 + \sum_{j=1}^{n-1}\beta_j\Delta ROA + \sum_{j=1}^{n-1}\gamma_j\Delta INF_{t-j} + \sum_{j=1}^{n-1}\rho_j\Delta MSP_{t-j} + \sum_{j=1}^{n-1}S_j\Delta EXCH_{t-j} + \sum_{j=1}^{n-1}H_j\Delta UPY_{t-j} + \varphi_2 ecm_{t-1} + \varepsilon_{2t} \quad\quad (3)$$

$$\Delta MSP_t = \alpha_1 + \sum_{j=1}^{n-1}\beta_j\Delta ROA + \sum_{j=1}^{n-1}\gamma_j\Delta INF_{t-j} + \sum_{j=1}^{n-1}\rho_j\Delta MSP_{t-j} + \sum_{j=1}^{n-1}S_j\Delta EXCH_{t-j} + \sum_{j=1}^{n-1}H_j\Delta UPY_{t-j} + \varphi_3 ecm_{t-1} + \varepsilon_{3t} \quad\quad (4)$$

$$\Delta EXCH_t = \alpha_1 + \sum_{j=1}^{n-1}\beta_j\Delta ROA + \sum_{j=1}^{n-1}\gamma_j\Delta INF_{t-j} + \sum_{j=1}^{n-1}\rho_j\Delta MSP_{t-j} + \sum_{j=1}^{n-1}S_j\Delta EXCH_{t-j} + \sum_{j=1}^{n-1}H_j\Delta UPY_{t-j} + \varphi_4 ecm_{t-1} + \varepsilon_{4t} \quad\quad (5)$$

$$\Delta UPY_t = \alpha_1 + \sum_{j=1}^{n-1}\beta_j\Delta ROA + \sum_{j=1}^{n-1}\gamma_j\Delta INF_{t-j} + \sum_{j=1}^{n-1}\rho_j\Delta MSP_{t-j} + \sum_{j=1}^{n-1}S_j\Delta EXCH_{t-j} + \sum_{j=1}^{n-1}H_j\Delta UPY_{t-j} + \varphi_5 ecm_{t-1} + \varepsilon_{5t} \quad\quad (6)$$

Where; t = 1, 2, 3 4, …T and j = 1, 2, 3, 4, …N, $ROA_t$ is the return on assets, $INF_t$ is the inflation rate, $MSP_t$ is money supply, $EXCH_t$ is the exchange rate, and $UPY_t$ is the unemployment rate. $\varepsilon_t$ is the error term. $ecm_{t-1}$ is the error correction term, while $\varphi_{3i}$ is the speed of adjustment if there is a deviation from the long-run estimator.

### Data and Estimation Techniques

This study is a secondary study that utilized time-series data collated in respect of the financial and macroeconomic variables. In this study, the proxy for macroeconomic dynamics is the unanticipated changes in the macroeconomic variables; exchange rate, inflation, money supply, and unemployment rate. The proxy for capital market performance is return on asset (ROA). The data were obtained from the Central Bank of Nigeria Statistical Bulletin and the World Development Indicator database (WDI). The estimation techniques adopted in this study include; the Phillips–Peron (1988) test of unit root, Johannsen test for cointegration, and vector error correction model (VECM). The VECM post estimation diagnostic tests utilized in this study are; the Lagrange multiplier (LM) test for autocorrelation in the residuals of vector error-correction models (VECMs), Jarque-Bera normality test for VECM, and the Eigenvalue stability condition check-in a vector error-correction model (VECM)

### IV. RESULTS AND DISCUSSION

### Unit Root Test

The unit root test adopted for this study is the Phillips–Peron (1988) test. The null hypothesis is that the variable contains a unit root; the alternative is that the variable is stationary. The result of the unit root test is in table 1. The result shows that all the variables are integrated of order one; they are all I (1) series. The result of the unit root test certified the conditions for applying the vector error correction model (VECM).

### Table 1: Unit Root Test

| Variable | Phillips-Peron Unit Root Test | |
|---|---|---|
| | Level | 1st Diff |
| INF | -8.156 | -22.416*** |
| EXR | -3.331 | -24.313*** |
| ROA | -4.175 | -29.188*** |
| UPY | -4.024 | -20.209*** |
| MSP | -0.635 | -14.812* |

*Source: Authors' computation (2021)*
Note: ***, **, *, denote statistical significance at the 1%, 5%, and 10% level respectively.

### Lag Length Selection

There is a need to determine the number of lags in the analysis to avoid multicollinearity in the study. The result of the lag selection analysis is in table 2. Three criteria selected three lags, and that is the minimum chosen by the criteria. Therefore, this study makes use of three lags.

### Table 2: Lag Length Selection

| Lag | df | P | FPE | AIC | HQIC | SBIC |
|---|---|---|---|---|---|---|
| 1 | 25 | | $7.0e^{-07}$ | 0.017846 | 0.076259 | 0.265811 |
| 2 | 25 | 0.000 | $10e^{-07}$ | -1.99713 | -1.64665 | -0.509347 |
| 3 | 25 | 0.000 | $4.8e^{-08}$ | -3.14739 | -2.50485 | -0.419788 |
| 4 | 25 | 0.000 | $4.7e^{-09}$ | -6.94069 | -6.00609 | -2.97327 |
| | 25 | 0.000 | 0* | | | |

*Source: Authors' computation (2021)*

### Cointegration Test

The Cointegration test for determining whether there is a level relationship among the variables or not is the Johannsen test for the cointegration approach. The result is in table 3. The result shows evidence of four cointegration equations in this study. Therefore, the study used the vector error correction model (VECM).

Table 3: Cointegration Test

| . vecrank ROA INF exch MSS UPY, trend (constant) lags(3) | | | | | |
|---|---|---|---|---|---|
| Johansen tests for cointergration | | | | | |
| Trend: constant  Sample: 1996 - 2017 | | | Number of obs = 22  Lags = 3 | | |
| Maximum rank | Parms | LL | Eigenvalue | Trace Statistic | 5% Critical Value |
| 0 | 55 | 129.71697 |  | 155.0606 | 68.52 |
| 1 | 64 | 162.46468 | 0.94906 | 89.5652 | 47.21 |
| 2 | 71 | 185.56569 | 0.87755 | 43.3632 | 29.68 |
| 3 | 76 | 199.38562 | 0.71531 | 15.7233 | 15.41 |
| 4 | 79 | 205.47901 | 0.42532 | 3.5365 ± | 3.76 |
| 5 | 80 | 207.24726 | 0.14850 |  |  |

**Source:** Authors' computation (2021)

### Capital Market Performance and Macroeconomic Dynamics in Nigeria

The effect of the macroeconomic dynamics in Nigeria on capital market performance is analysed using the vector error correction model (VECM). The result of the analysis when the return on asset (ROA) is the dependent variable is in table 4. The cointegration term, which is the speed of adjustment, is estimated at '-0.185603'. The coefficient of the speed of adjustment is negatively signed, which implies a convergence among the variables in the long-run. The result suggests that disequilibrium from the previous year's shock converges back to the long-run equilibrium in the current year at the speed of 18%. This further implies long-run causality running from the exchange rate, inflation, money supply, and unemployment rate to capital market performance in Nigeria.

The result supports the Arbitrage Pricing Theory (APT) proposition, which stipulates that asset returns is forecastable through the linear relationship between an asset's expected returns and the macroeconomic factors whose dynamics affect the asset's risk. In other words, the result of this study supports the proposition that the dynamics in the exchange rate, inflation, money supply, and unemployment rate influence the capital market performance in Nigeria.

The short-run analysis reveals the existence of short-run causality running from inflation, money supply, and unemployment rate to ROA; with the exception of the exchange rate. There is also a short-run causality running from the one period lag of the ROA to its current value. In other words, the one-year history of the ROA influences its current status in the short run.

The result of the analysis when the exchange rate is the dependent variable is in the second column of Table 4. When the exchange rate is the dependent variable, the result shows long-run causality running from return on assets, inflation, money supply, and unemployment rate. In other words, the speed of adjustment, which determines the long-run relationship, is estimated at '-0.5209245'; the coefficient of the speed of adjustment is negatively signed and significant. The result implies that there is convergence among the variables in the long-run. A disequilibrium from the previous year's shock converges back to the long-run equilibrium in the current year at the speed of 52%. This result and the previous one signify among other things that causality does not run from the exchange rate to the capital market performance indicator (ROA). Instead, causality runs from the capital market performance indicator (ROA) to the exchange rate. The result of the short-run analysis shows that short-run causality only runs from inflation rate to exchange rate. But the effect of the inflationary shock on the exchange rate will not endure for an extended period because the speed of adjustment is above 50%.

Table 4: Result of the Analysis When ROA, EXCH, INF, MSS, and UPY are Dependent Variable Respectively

| Explanatory Variables \ Dependent Variable | ROA | EXCH | INF | MSS | REPLY |
|---|---|---|---|---|---|
| | Coefficients | Coefficients | Coefficients | Coefficients | Coefficients |
| ECM | -0.18560*** | 0.520924* | -0.32615 | -0.010377 | -0.06985 |
| ROA$_{t-1}$ | -0.69831*** | -0.003443 | -0.15249 | -0.54012 | -0.02276 |
| ROA$_{t-2}$ | -0.12437 | -0.000081 | -0.19311 | -0.00526 | -0.03736 |
| EXCH$_{t-1}$ | 11.57623 | 0.2570836 | 6.407513 | 0.015204 | 0.63844 |
| EXCH$_{t-2}$ | 23.49526 | -0.148972 | -20.25355 | 0.706230 | 0.58876 |
| INF$_{t-1}$ | -2.34515*** | -0.014138* | 0.152313 | -0.118599 | -0.03399 |
| INF$_{t-2}$ | -1.37211*** | 0.0023728 | 0.081077 | -0.066275 | -0.74065 |
| MSS$_{t-1}$ | 2.40742** | 0.01074 | 1.46859 | 0.864549* | -0.06055 |
| MSS$_{t-2}$ | 0.10168 | 0.014563 | -0.39944 | -0.426590 | 0.06326 |
| UPY$_{t-1}$ | 1.60029* | 0.002584 | -1.47792 | -0.68801 | -0.15654 |
| UPY$_{t-2}$ | 3.44158 | 0.038128 | -3.548768 | -0.41746 | -2.77927*** |

Note: ***, **, *, denote statistical significance at the 1%, 5%, and 10% level respectively.
**Source:** *Authors' computation (2021)*

When the inflation rate is the dependent variable, all the short-run coefficients and the coefficient of the error correction term are not statistically significant. The result implies no long-run or short-run causality running from return on assets, money supply, exchange rate, and the unemployment rate to inflation rate in Nigeria. In other words, factors beyond those macroeconomic variables propel the inflationary pressure in Nigeria.

The analysis result when money supply is the dependent variable is in the third column of Table 4. The coefficient of the error correction term is negative and not significant. This result implies long-run causality running from return on assets, inflation, exchange rate, and the unemployment rate to the money supply in Nigeria. However, short-run causality runs from the one-period lag of money supply and the unemployment rate to the current money supply in Nigeria.

When the unemployment rate is the dependent variable, there is no long-run causality running from return on assets, inflation, money supply, and exchange rate to the unemployment rate in Nigeria. The result also shows no significant short-run relationship between the variables (return on assets, inflation, money supply, and exchange rate) and the unemployment rate in Nigeria. It implies that the unemployment rate in Nigeria is not determined by any of those macroeconomic factors. The unemployment problem in Nigeria is beyond macroeconomic manipulation. It is a systemic problem that only an employment-generation oriented structural change can tackle.

### Diagnostic Tests

The diagnostic tests that associate with VECM include the Lagrange multiplier (LM) test for autocorrelation in the residuals of vector error-correction models (VECMs), Jarque-Bera normality test for VECM, and the Eigenvalue stability condition check-in a vector error-correction model (VECM). The result of the autocorrelation test is in table 5. The result shows that the analysis is free from autocorrelation problems up to two lags. The result shows that there is an autocorrelation problem at the third lag. In other words, the null hypothesis of no autocorrelation is accepted at the first two lags. At the third lag, the probability is significant at a 5% level of significance, which implies the presence of autocorrelation in the model starting from the third lag.

Table 5: Autocorrelation Test

| .veclmar, mlag (3) | | | |
|---|---|---|---|
| Lagrange-multiplier test | | | |
| Lag | chi2 | df | Prob > chi2 |
| 1 | 15.5313 | 25 | 0.92807 |
| 2 | 33.6174 | 25 | 0.11634 |
| 3 | 51.8166 | 25 | 0.00126 |

H0: no autocorrelation at lag order
**Source:** *Authors' computation (2021)*

The result of the normality test is in table 6. The null hypothesis is that the disturbances in a VECM are normally distributed. Therefore, any estimate whose probability is higher than the 10% level of significance is regarded as being normally distributed. While the ones that are significant at 1%, 5%, or 10% level of significance are not normally distributed. Therefore, in this analysis, only money supply is not normally distributed. Other variables in the study are normally distributed.

Table 6: Normality Test

| Jarque-Bera test | | | |
|---|---|---|---|
| Equation | chi2 | df | Prob > chi2 |
| D_UPY | 3.698 | 2 | 0.15743 |
| D_MSS | 31.944 | 2 | 0.00000 |
| D_INF | 0.532 | 2 | 0.76628 |
| D_EXCH | 2.365 | 2 | 0.30650 |
| D_ROA | 0.034 | 2 | 0.98331 |
| ALL | 38.573 | 10 | 0.00003 |

*Source:* Authors' computation (2021)

The VECM analysis is based on the evidence of a cointegrating relationship among the endogenous variables in the model. The reliability of a VECM result is conditioned on whether the cointegrating equations are stationary or not and whether the number of cointegrating equations are correctly specified or not. The Eigenvalues Stability test usually indicates whether the number of cointegrating equations is miss-specified or whether the cointegrating equations, which are assumed to be stationary, are not stationary. The decision rule is that the companion matrix of a VECM with $P$ endogenous variables and $S$ cointegrating equations has $P - S$ unit Eigenvalues. If the process is stable, the moduli of the remaining $S$ eigenvalues are strictly less than one. In this study, $P=5$ and $S=4$. The result in table 7 shows that the cointegrating equations in this analysis are stable. The moduli of the eigenvalues assumed the values from 0.86 to 0.37, which are less than one, implying that the model is stable.

Table 7: Stability Test

| . vecstable | |
|---|---|
| Eigenvalue stability condition | |
| Eigenvalue | Modulus |
| -.1130194 + 1.66313i | 1.66697 |
| -.1130194 – 1.66313i | 1.66697 |
| 1 | 1 |
| 1 | 1 |
| 1 | 1 |
| 1 | 1 |
| .6111285 + .6187434i | .869667 |
| .6111285 - .6187434i | .869667 |
| -.7936705 | .793671 |
| .4164019 + .5775689i | .712023 |
| .4164019 - .5775689i | .712023 |
| -.3931113 + .5878555i | .707185 |
| -.3931113 - .5878555i | .707185 |
| .02852327 + .3748152i | .375899 |
| .02852327 - .3748152i | .375899 |

The VECM specification imposes 4 unit moduli.
*Source:* Authors' computation (2021)

## V. CONCLUSION

This study examined the relationship between capital market performance and the macroeconomic dynamics in Nigeria. The result obtained from the analysis revealed a significant long-run relationship between capital market performance and the macroeconomic dynamics in Nigeria. We observed long-run causality running from the exchange rate, inflation, money supply, and unemployment rate to capital market performance in Nigeria. The result supports the Arbitrage Pricing Theory (APT) proposition in the Nigerian context. The theory stipulates that asset returns can be forecasted through the linear relationship between an assets's expected returns and the macroeconomic factors whose dynamics affect the asset's risk. In other words, the result of this study supports the proposition that the dynamics in the exchange rate, inflation, money supply, and unemployment rate influence the capital market performance in Nigeria. The study validates the proposition of Arbitrage Pricing Theory (APT) in Nigeria.

## References

Acha, I., & Akpan, S. (2019). Capital market performance and economic growth in Nigeria. *Nobel International Journal of Economics and Financial Research*, *4*(2), 10–18.

Alam, M. S., & Hussein, M. A. (2019). The impact of capital market on the economic growth in Oman. *"Victor Slăvescu" Centre for Financial and Monetary Research, Bucharest*, *(23)*, 2(84), 117-129. Romanian Academy, National Institute of Economic Research (INCE), Financial Studies, ISSN 2066-6071,

Anigbogu, U., & Nduka, E. (2014). Stock market performance and economic growth: evidence from Nigeria employing vector error correction model. *The Economics and Finance Letters*, *1*, 90-103.

Ankargren, S., Bjellerup, M., & Shahnazarian, H. (2017). The importance of the financial system for the real economy. *Empirical Economics*, *53*(4), 1553-1586.

Cynthia, U. G., Chinedum, N. N., & Ikechi, K. S. (2021). Effects of capital market development on the economic growth of Nigeria. *International Journal of Innovation and Economic Development*

Edame, G. E., & Okoro, U. (2013). The impact of capital market on economic growth in Nigeria. *Journal of Poverty, Investment and Development*, *(1)*1.

Enoruwa, K. O., Ezuem, M. D., & Nwani, O. C. (2019). Capital market performance indicators and economic growth in Nigeria. *International Journal of Research and Innovation in Social Science*, *3*(2), 435-444.

Okoye, V. O., & Nwisienyi, K.J. (2013). The capital market contributions towards economic growth and development; the Nigerian experience. *Global Advanced Research Journal of Management and Business Studies*, *(2)*2. Available online at http://garj.org/garjmbs/index.htm

Odetayo, T. A., & Sajuyigbe, A. S. (2012). Impact of Nigerian capital market on economic growth and development. *International Journal of Arts and Commerce*, *1*(5).

Okodua, H., & Ewetan, O. (2013). Stock market performance and sustainable economic growth in Nigeria: A bounds testing co-integration approach. *Journal of Sustainable Development*, *6*.

Phillips, P. C. B., & Perron, P. (1988). Testing for Unit Roots in Time Series Regression. Biometrika, 75, 335-346

Ross, S. A. (1976). The arbitrage theory of capital asset pricing. *Journal of Economic Theory*. *13*(3), 341–360. doi:10.1016/0022-0531(76)90046-6. ISSN 0022-0531.

Shallu. (2014). Indian capital market and impact of SEBI. *Tactful Management Research Journal*, *2*(4), 1-10.